# New Beam Dynamics Code for Cyclotron Analysis

G-H. Kim, [1] H-J. Cho, [1] B-H. Oh,[1,+] G-R. Hahn,[2] M. Chung,[3] S. Park, [1] and S. Shin[1,*]

[1] Department of Accelerator Science, Korea University, Sejong, 30019, KOREA
[2] Pohang Accelerator Laboratory, Pohang, Kyungbuk, 37673, KOREA
[3] Ulsan National Institute of Science and Technology, Ulsan, 44919, KOREA

E-mail:[*] tlssh@korea.ac.kr, [+] obh96@korea.ac.kr

**Abstract**. This paper describes the beam dynamic simulation with transfer matrix method for cyclotron. Starting from a description on the equation of motion in the cyclotron, lattice functions were determined from transfer matrix method and the solutions for the 2$^{nd}$-order nonlinear Hamiltonian were introduced and used in phase space particle tracking. Based on the description of beam dynamics in the cyclotron, simulation code was also developed for cyclotron design.

## 1. Introduction

Although the particle energy that can be obtained from a cyclotron is limited and cannot be expanded to high energy particle accelerator, its utilization is still increasing in various fields such as medical, industrial, and research applications. In addition, MW class high power cyclotron [1, 2] are also being actively studied for some applications; the production of neutrons in a spallation source, the production of muons via pions from a fixed target, the production of neutrinos for particle physics applications and accelerator driven system (ADS) [3].

As in other accelerator design and development, beam dynamics calculation programs are essential in cyclotron design and development for various applications. In particular, in the case of high-power cyclotron, new concept designs are being proposed to solve challenging beam physics problems, and the beam dynamics calculation programs for cyclotron must be advanced accordingly. For example, next generation strong-focusing cyclotron [4] suggests using three stacked cyclotrons, each one designed to accelerate more than 10 mA. To achieve this ambitious goal, a set of 35 quadrupole focusing channels are installed in each hill of each cyclotron. In this design, nonlinear effects and chromatic effects from strong quadrupole are large, so existing beam dynamics calculation programs are limited. This invokes the need for new beam dynamics program development for cyclotron analysis.

Beam dynamics calculations in a cyclotron can be performed in two methods: 4th-order Runge-Kutta and Equilibrium Orbit (EO). 4th-order Runge-Kutta method directly and numerically solves general Lorenz's equation using three dimensional (3D) field map. This method uses a 3D field map to directly track particles and then indirectly calculate beam properties. However, it takes a relatively long time to obtain solution and has limitations in gaining direct insight from cyclotron design perspective, so it is not suitable for design process that requires many iterations and it ultimately used for verification simulation after design completion. In contrast, the EO method constructs a canonical equation of motion in cyclotron based on theory and then solves the canonical equation directly numerically using only the median plane field to find EO at a given particle energy. Once EO is found, the related properties such as linear radial and vertical tunes and phase excursion can be calculated immediately, making it effective and useful when designing a cyclotron.

These days, with the trend of using artificial intelligence to optimize accelerator design, beam dynamics program using EO method that take a short time to calculate beam parameters are increasingly needed for cyclotron design. However, as previously mentioned, improvement of the existing EO method program is necessary in designing a new concept cyclotron. Rather than describing beam motion using the lattice function including dispersion function and dealing beam motion with nonlinear effect as in general beam dynamics program, beam dynamics program using EO method calculates the closed orbit at given energy and then computes the corresponding beam parameters, such as tune and isochronous. Therefore, starting from the closed orbit calculation of the existing EO method, new beam dynamics program had been development. In new beam dynamics program, the lattice function including dispersion function was obtained to effectively describe beam motion, and 2nd-order transfer matrix was solved to describe nonlinear motion in canonical variable space. In this paper, based on the theoretical exploration on cyclotron beam dynamics, we introduce the developed program for designing advanced cyclotron. Section 2 derives the second order equation of motion in cyclotron and Section 3 describes determination of the lattice function including dispersion function. Section 4 gives results from 2nd-order transfer-matrix form Hamiltonian. Finally, Section 5 presents the conclusion and summary.

## 2. Hamiltonian in Cyclotron

In the context of a cyclotron, which is a type of particle accelerator distinguished by its mirror symmetry around a mid-plane, we begin our exploration by considering the magnetic field expressed in polar coordinates. This is denoted by Eq. (1), which describes the behavior of the magnetic field within the cyclotron.

$$B_\theta(r,\theta,0) = 0,$$
$$B_r(r,\theta,z) = -B_r(r,\theta,-z),$$
$$B_\theta(r,\theta,z) = -B_\theta(r,\theta,-z),$$
$$B_z(r,\theta,z) = B_z(r,\theta,-z).$$
(1)

In a cyclotron, we can describe the magnetic field in three dimensions as a function of z, given that $z/r \ll 1$, and the conditions $\vec{\nabla} \cdot \vec{B} = \vec{\nabla} \times \vec{B} = 0$ are met. We can simplify the expression by ignoring terms beyond the second order. This simplification results in the following expansion from [7,8]

$$B_r = z\frac{\partial B}{\partial r},$$
$$B_\theta = \frac{z}{r}\frac{\partial B}{\partial \theta},$$
$$B_z = B - \frac{1}{2}z^2\left(\frac{\partial^2 B}{\partial r^2} + \frac{1}{r}\frac{\partial B}{\partial r} + \frac{1}{r^2}\frac{\partial^2 B}{\partial \theta^2}\right),$$
(2)

Suppose there is a charged particle with momentum $\vec{p}$ and charge $q$ moving in a magnetic field with vector potential $\vec{A}$. In this case, the standard momentum of the particle $\vec{P}$ can be expressed by equation (3). And the action integral S can be defined using the cylindrical coordinates through equations (4), respectively.

$$\vec{P} = \vec{p} + q\vec{A},$$
(3)

$$S = \int_{t_0}^{t}(P_x\frac{dx}{dt} + P_y\frac{dy}{dt} + P_z\frac{dz}{dt} - \mathcal{H})dt$$
$$= \int_{t_0}^{t}\left(P_r\frac{dr}{d\theta} + P_\theta\frac{d\theta}{d\theta} + P_z\frac{dz}{d\theta} - \mathcal{H}\frac{dt}{d\theta}\right)d\theta = \int_{t_0}^{t}\left(P_r\frac{dr}{d\theta} + P_z\frac{dz}{d\theta} - \mathcal{H}\frac{dt}{d\theta} + P_\theta\right)d\theta.$$
(4)

In equation (4), if we select angle θ instead of time t as the independent variable, we can obtain new Hamiltonian conjugate variables as explained in reference [9].

$$-E \ (= H) \ ; \ t,$$
$$P_\theta = rp_\theta + qrA_\theta \ ; \ \theta,$$
$$P_r = p_r + qA_r \ ; \ r,$$
$$P_z = p_z + qA_z \ ; \ z$$
(5)

and the new Hamiltonian is

$$H = -P_\theta = -rp_\theta - qrA_\theta = -r\sqrt{(P^2 - (P_r - qA_r)^2 - (P_z - qA_z)^2)} - qrA_\theta.$$
(6)

The equation $\tau = \omega_0 t$ represents the relationship between revolution frequency $\omega_0$ and the travel time of particles $t$. By applying a magnetic field divided by charge $q$ (i.e., $\vec{B} \to \vec{B}/q$), we can solve the Hamiltonian equation with the new Hamiltonian of equation (6) referred to in [8] to derive the equation of motion.

$$\frac{dr}{d\theta} = \frac{\partial H}{\partial P_r} = \frac{r(P_r - A_r)}{\sqrt{(P^2 - (P_r - A_r)^2 - (P_z - A_z)^2)}} = \frac{rp_r}{\sqrt{(P^2 - p_r^2 - p_z^2)}} = \frac{rp_r}{p_\theta},$$

$$\frac{dz}{d\theta} = \frac{\partial H}{\partial P_z} = \frac{r(P_z - A_z)}{\sqrt{(P^2 - (P_r - A_r)^2 - (P_z - A_z)^2)}} = \frac{rp_z}{\sqrt{(P^2 - p_r^2 - p_z^2)}} = \frac{rp_z}{p_\theta},$$

$$\frac{dp_r}{d\theta} = -\frac{\partial H}{\partial r} - \frac{\partial A_r}{\partial r}\frac{dr}{d\theta} - \frac{\partial A_r}{\partial z}\frac{dz}{d\theta} - \frac{\partial A_r}{\partial \theta} = p_\theta - r\left(B - \frac{1}{2}z^2(\vec{\nabla}_\perp^2 B)\right) + \frac{zp_z}{p_\theta}\frac{\partial B}{\partial \theta},$$

$$\frac{dp_z}{d\theta} = -\frac{\partial H}{\partial z} - \frac{\partial A_z}{\partial r}\frac{dr}{d\theta} - \frac{\partial A_z}{\partial z}\frac{dz}{d\theta} - \frac{\partial A_z}{\partial \theta} = z\left(r\frac{\partial B}{\partial r} - \frac{dr/d\theta}{r}\frac{\partial B}{\partial \theta}\right),$$

$$\frac{d\tau}{d\theta} = -\omega_0 \frac{\partial H}{\partial E} = \frac{\omega_0 \frac{rE}{c^2}}{\sqrt{\left(\frac{E^2}{c^2} - m^2c^2 - (P_r - eA_r)^2 - (P_z - eA_z)^2\right)}} = \frac{\omega_0 \gamma m r}{p_\theta} = \frac{\gamma r}{\frac{p_\theta}{m\omega_0}}, \quad (7)$$

The equation for calculating the total energy (E) of an object is derived using the object's total mass (m), the relativistic factor (γ), and the velocity of light (c). When analyzing the total energy equation, both the values of z and p_z are considered to be zero. By applying this to equation (7), the equation for total energy (EO) can be expressed in a simplified form.

$$\frac{dr}{d\theta} = \frac{rp_r}{p_\theta}.$$
$$\frac{dp_r}{d\theta} = p_\theta - rB,$$
$$\frac{d\tau}{d\theta} = \frac{\gamma r}{\frac{p_\theta}{m\omega_0}}. \quad (8)$$

In the study of a system's coordinates, a technique of expanding the values of r, p_r, and τ by adding x, p_x, and χ respectively is employed. This expansion technique enables the discovery of the equations of displaced orbit, which reveal the deviation of a particle's orbit from the initial values of the system. To find these initial values, an efficient iteration process, similar to the Newton method, is utilized. By substituting r, p_r, and τ with their expanded terms r+x, p_r+p_x, and τ+ χ in the equation of motion (Eq. 7), the equations of displaced orbit can be derived. Moreover, these equations are further expanded in terms of x, p_x, z, p_z, and τ up to 2nd order, resulting in 2nd-order equations of motion. This process provides a comprehensive understanding of the system's coordinates and enables the prediction of the deviation of a particle's orbit from its initial values.

$$\frac{dx}{d\theta} = \frac{p_r}{p_\theta}x + \frac{rP^2}{p_\theta^3}p_x + \frac{P^2}{p_\theta^3}xp_x + \frac{3}{2}\frac{rp_rP^2}{p_\theta^5}p_x^2 + \frac{1}{2}\frac{rp_r}{p_\theta^3}p_z^2,$$

$$\frac{dp_x}{d\theta} = -\frac{p_r}{p_\theta}p_x - \left(B + r\frac{\partial B}{\partial r}\right)x - \frac{1}{2}\frac{P^2}{p_\theta^3}p_x^2 - \frac{1}{2}\left(2\frac{\partial B}{\partial r} + r\frac{\partial^2 B}{\partial x^2}\right)x^2 - \frac{1}{2}\frac{1}{p_\theta}p_z^2 + \frac{1}{2}r(\vec{\nabla}^2 B)z^2 + \frac{\partial B}{\partial \theta}\frac{1}{p_\theta}zp_z,$$

$$\frac{dz}{d\theta} = \frac{r}{p_\theta}p_z + \frac{1}{p_\theta}p_z x + \frac{rp_r}{p_\theta^3}p_x p_z,$$

$$\frac{dp_z}{d\theta} = \left(r\frac{\partial B}{\partial r} - \frac{p_r}{p_\theta}\frac{\partial B}{\partial \theta}\right)z + \left(\frac{\partial B}{\partial r} + r\frac{\partial^2 B}{\partial r^2} - \frac{p_r}{p_\theta}\frac{\partial^2 B}{\partial r\partial \theta}\right)xz - \frac{p_r^2}{p_\theta^3}\frac{\partial B}{\partial \theta}zp_x,$$

$$\frac{d\tau}{d\theta} = m\omega_0\left[\frac{\gamma r}{p_\theta}x + \frac{\gamma r p_r}{p_\theta^3}p_x + \frac{\gamma p_r}{p_\theta^3}xp_x + \frac{1}{2}\left(\frac{\gamma r}{p_\theta^3} + \frac{3\gamma r p_r^2}{p_\theta^5}\right)p_x^2 + \frac{1}{2}\frac{\gamma r}{p_\theta^3}p_z^2\right] \quad (9)$$

By neglecting all nonlinear terms in Eqs. (9), a linear equation of motion can be expressed as

$$\frac{dx}{d\theta} = \frac{p_r}{p_\theta}x + \frac{rP^2}{p_\theta^3}p_x,$$
$$\frac{dp_x}{d\theta} = -\frac{p_r}{p_\theta}p_x - \left(B + r\frac{\partial B}{\partial r}\right)x,$$
$$\frac{dz}{d\theta} = \frac{r}{p_\theta}p_z,$$
$$\frac{dp_z}{d\theta} = \left(r\frac{\partial B}{\partial r} - \frac{p_r}{p_\theta}\frac{\partial B}{\partial \theta}\right)z. \tag{10}$$

Solutions of equation Eq. (10) can be written in matrix form

$$\begin{pmatrix} x \\ p_x \end{pmatrix} = X\begin{pmatrix} x_0 \\ p_{x0} \end{pmatrix} = \begin{pmatrix} X_{11}(\theta) & X_{12}(\theta) \\ X_{21}(\theta) & X_{22}(\theta) \end{pmatrix}\begin{pmatrix} x_0 \\ p_{x0} \end{pmatrix},$$
$$\begin{pmatrix} z \\ p_z \end{pmatrix} = Z\begin{pmatrix} z_0 \\ p_{z0} \end{pmatrix} = \begin{pmatrix} Z_{11}(\theta) & Z_{12}(\theta) \\ Z_{21}(\theta) & Z_{22}(\theta) \end{pmatrix}\begin{pmatrix} z_0 \\ p_{z0} \end{pmatrix} \tag{11}$$

Matrices X and Z can be calculated using numerical methods such as the Runge-Kutta method.

## 3. Lattice functions in Cyclotron

This Session describes the calculations of several essential beam parameters implemented in the new beam dynamics code for cyclotron analysis. Components in the transfer matrices can be calculated by numerical methods when the magnetic field distribution is known. The field distribution (Fig. 1) from reference [11] was used. Two important parameters in beam optics are the radial and the vertical focusing frequencies, which are defined as the number of betatron oscillations per revolution of a reference particle. These frequencies measure the degree of focusing in the radial and the vertical planes, respectively. Once an EO is found, the linearized equations of motion are integrated for one sector (or one turn) to compute the transfer matrices. The focusing frequencies $v_r$ and $v_z$ are then computed using

$$\cos(2\pi v_r) = \frac{1}{2}Tr(M_r),$$
$$\cos(2\pi v_z) = \frac{1}{2}Tr(M_z), \tag{12}$$

where $M_r$ and $M_z$ are the transfer matrices for one turn and the symbol $Tr$ signifies the trace of the matrix [12]. The magnetic fields must be carefully designed to avoid harmful resonances during the entire acceleration process. For the four-sector cyclotron that we consider here, the following structure resonances must be considered to affect the beam:

$$4v_r = 4$$
$$3v_z = 4$$
$$2v_z = 4$$
$$v_r - 2v_z = 0$$
$$2v_r + 2v_z = 4. \tag{13}$$

The particle's focusing frequencies diverge between injection and extraction (Fig. 2).

During a transformation along a circle in a cyclotron, the phase ellipse will continuously change its form and orientation, but not its area at a given energy. In matrix formulation, the ellipse parameters, which are also called Twiss parameters, the transform from initial Twiss parameters is given by [12]

$$\begin{pmatrix} \alpha \\ \beta \\ \gamma \end{pmatrix}_\theta = \begin{pmatrix} Y_{11}^2 & -2Y_{11}Y_{12} & Y_{12}^2 \\ -Y_{11}Y_{21} & Y_{11}Y_{21} + Y_{12}Y_{21} & -Y_{12}Y_{22} \\ Y_{21}^2 & -2Y_{21}Y_{22} & Y_{22}^2 \end{pmatrix}\begin{pmatrix} \alpha \\ \beta \\ \gamma \end{pmatrix}_{\theta_i} \tag{14}$$

where element $Y_{ij}$ of the of transfer matrix indicates $X_{ij}$ for the radial plane and $Z_{ij}$ for the vertical

plane. Generally, the particle trajectory and its derivative in Twiss parameters can be expressed as [12]

$$\begin{pmatrix} y \\ p_y \end{pmatrix}_\theta = \begin{pmatrix} \sqrt{\frac{\beta}{\beta_0}}(\cos\mu + \alpha_0 \sin\mu) & \sqrt{\beta\beta_0}\sin\mu \\ -\frac{(\alpha-\alpha_0)\cos\mu + (1+\alpha\alpha_0)\sin\mu}{\sqrt{\beta\beta_0}} & \sqrt{\frac{\beta_0}{\beta}}(\cos\mu - \alpha\sin\mu) \end{pmatrix} \begin{pmatrix} y \\ p_y \end{pmatrix}_{\theta_i}$$

(15)

If the cyclotron is a circular accelerator, a periodic condition can be imposed with ($\alpha = \alpha_0$, $\beta = \beta_0$) after one turn, and the transfer matrix is given by

$$\begin{pmatrix} Y_{11} & Y_{21} \\ Y_{12} & Y_{22} \end{pmatrix} = \begin{pmatrix} \cos\mu + \alpha\sin\mu & \beta\sin\mu \\ -\gamma\sin\mu & \cos\mu - \alpha\sin\mu \end{pmatrix}$$

(16)

Here phase term $\mu = 2\pi\nu$ at $\theta = 2\pi$. The initial Twiss function can be determined when an element of the transfer-matrix (Eq. 11) and focusing frequencies (Eq. 12) are calculated numerically. Beta-functions (Fig. 3) determined from the transfer matrix in Eq. (14) have useful information during magnet design for cyclotron by implying beam size variation from initially assumed beam. In the figure, beta-function shows four-fold symmetry because the cyclotron has four sectors.

A cyclotron exerts no phase-restore force, so energy gain accumulates until extraction. Especially, a space-charge force can dramatically distort the beam distribution in longitudinal phase space in cyclotron. Therefore, the dispersion function is an important parameter in descriptions of beam dynamics in cyclotrons, because it indicates longitudinally-dependent radial motion. This function in a cyclotron is

$$\eta(E) = \frac{Xco(E) - Xco(E+\Delta E)}{\delta}$$

(17)

where Xco is a closed orbit at given energy, and $\delta$ is momentum deviation between a given energy and the next energy. This dispersion function (Fig. 4) also shows four-fold symmetry due to the four-sector cyclotron. These functions can be useful in determining beam dynamics for exotic cyclotron [4].

### 4. 2nd order transfer-matrix in Cyclotron

In the second section of this paper, the author derived the second order equations of motion (as shown in Equation (9)) and provided solutions to the corresponding linear equation in matrix form (as shown in Equation (11)). By substituting the first-order solutions (also given in Equation (11)) into the nonlinear terms of Equation (9), the nonlinear equations were transformed into linear inhomogeneous equations. This involved defining f, g, and h (as described in Table 1). The approximate solutions can be obtained by solving the resulting inhomogeneous equations.

$$\frac{dx}{d\theta} = \frac{p_r}{p_\theta}x + \frac{rP^2}{p_\theta^3}p_x + f_{11}x_0^2 + f_{12}x_0 p_{x0} + f_{13}p_{x0}^2 + f_{14}z_0^2 + f_{15}z_0 p_{z0} + f_{16}p_{z0}^2$$

$$\frac{dp_x}{d\theta} = -\frac{p_r}{p_\theta}p_x - \left(B + r\frac{\partial B}{\partial r}\right)x + f_{21}x_0^2 + f_{22}x_0 p_{x0} + f_{23}p_{x0}^2 + f_{24}z_0^2 + f_{25}z_0 p_{z0} + f_{26}p_{z0}^2$$

$$\frac{dz}{d\theta} = \frac{r}{p_\theta}p_z + g_{11}x_0 z_0 + g_{12}x_0 p_{z0} + g_{13}p_{x0}z_0 + g_{14}p_{x0}p_{z0},$$

$$\frac{dp_z}{d\theta} = \left(r\frac{\partial B}{\partial r} - \frac{p_r}{p_\theta}\frac{\partial B}{\partial \theta}\right)z + g_{21}x_0 z_0 + g_{22}x_0 p_{z0} + g_{23}p_{x0}z_0 + g_{24}p_{x0}p_{z0}$$

$$\frac{d\tau}{d\theta} = \frac{\gamma r}{p_\theta}x + \frac{\gamma r p_r}{p_\theta^3}p_x + h_1 x_0^2 + h_2 x_0 p_{x0} + h_3 p_{x0}^2 + h_4 z_0^2 + h_5 z_0 p_{z0} + h_6 p_{z0}^2 \quad (18)$$

If Green's function is defined as

$$\Lambda_x(\theta,\theta') = X(\theta)X(\theta')$$

$$\Lambda_z(\theta, \theta')) = Z(\theta)Z(\theta'), \tag{19}$$

where X and Z are matrixes in Eq. (11) and vectors $V_x$ and $V_z$ are given as

$$V_x = \begin{pmatrix} x_0^2 \\ x_0 p_{x0} \\ p_{x0}^2 \\ z_0^2 \\ z_0 p_{z0} \\ p_{z0}^2 \end{pmatrix}, \quad V_z = \begin{pmatrix} x_0 z_0 \\ x_0 p_{z0} \\ p_{x0} z_0 \\ p_{x0} p_{z0} \end{pmatrix}. \tag{20}$$

then solutions of Eq. (18) are expressed as

$$\begin{pmatrix} x \\ p_x \end{pmatrix} = \begin{pmatrix} X_{11}(\theta) & X_{12}(\theta) \\ X_{21}(\theta) & X_{22}(\theta) \end{pmatrix} \begin{pmatrix} x_0 \\ p_{x0} \end{pmatrix} + \sum_{n=1}^{6} \left( \int_0^\theta \Lambda_x(\theta, \theta') \begin{pmatrix} f_{1i}(\theta') \\ f_{2i}(\theta') \end{pmatrix} d\theta' \right) V_{x,n},$$

$$\begin{pmatrix} z \\ p_z \end{pmatrix} = \begin{pmatrix} Z_{11}(\theta) & Z_{12}(\theta) \\ Z_{21}(\theta) & Z_{22}(\theta) \end{pmatrix} \begin{pmatrix} z_0 \\ p_{z0} \end{pmatrix} + \sum_{n=1}^{4} \left( \int_0^\theta \Lambda_z(\theta, \theta') \begin{pmatrix} g_{1i}(\theta') \\ g_{2i}(\theta') \end{pmatrix} d\theta' \right) V_{z,n}. \tag{21}$$

Similarly, solution $\tau$ can be described as

$$\tau = \int_0^\theta \left( \frac{\gamma r}{p_\theta}(X_{11}(\theta)x_0 + X_{12}(\theta)p_{x0}) + \frac{\gamma r p_r}{p_\theta^3}(X_{21}(\theta)x_0 + X_{22}(\theta)p_{x0}) \right) d\theta'$$
$$+ \sum_{i=1}^{6} \left[ \int_0^\theta \left( \frac{\gamma r}{p_\theta} a_{1n} + \frac{\gamma r p_r}{p_\theta^3} a_{2n} + h_n \right) d\theta' V_{x,n} \right] \tag{22}$$

The study involved the tracking of particles within a 13-MeV cyclotron. To generate phase-space ellipses, three different methods were employed: the linear transfer-matrix method (Eq. 11), the 2nd-order transfer-matrix method (Eq. 21), and the full Runge-Kutta method (Eq. 9). The research revealed that the 2nd-order transfer-matrix method and the full Runge-Kutta method produced better agreement than the linear transfer method and the full Runge-Kutta method. It is essential to note that the Runge-Kutta method is relatively slower than the other two methods. Figure 5 provides a visual representation of the generated phase-space ellipses, which demonstrate the characteristics of the particle trajectories.

## 5. Conclusion
This paper has described development of new beam dynamics code for the next generation strong-focusing cyclotron. Lattice functions, which are useful parameters to describe beam dynamics in circular accelerator, were calculated from a transfer matrix for the first time in cyclotron beam dynamics. In addition, the 2$^{nd}$-order transfer matrix was explored, and the phase-space beam motion from the 2$^{nd}$-order transfer matrix was compared with the result by Rounge-Kutta method. The phase-space ellipses from the 2$^{nd}$-order transfer-matrix method agreed well with the result obtained using the full Runge-Kutta method. Based on these beam dynamics results, cyclotron simulation code has been developed; it will be available on the request.

**Acknowledgments**
We thank to M. Yoon (POSTECH) for useful discussions. This work was supported by a Korea University Grant and also by Basic Science Research Program through the National Research Foundation of Korea (NRF-2019R1A2C1004862).

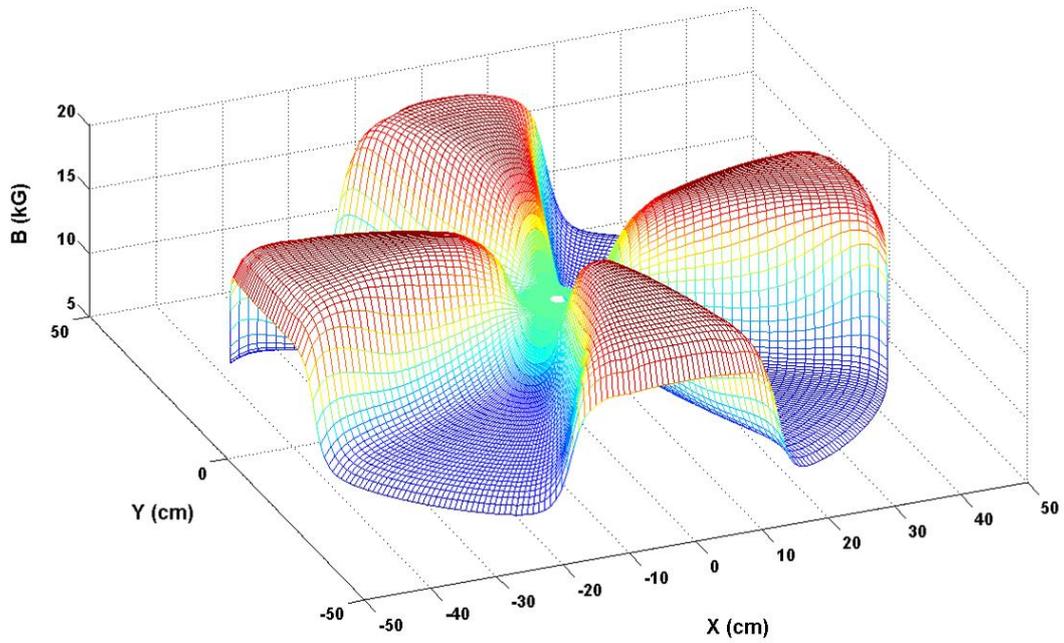

**Figure 1.** Field distribution for sector focused cyclotron [11]. Here sector number is 4.

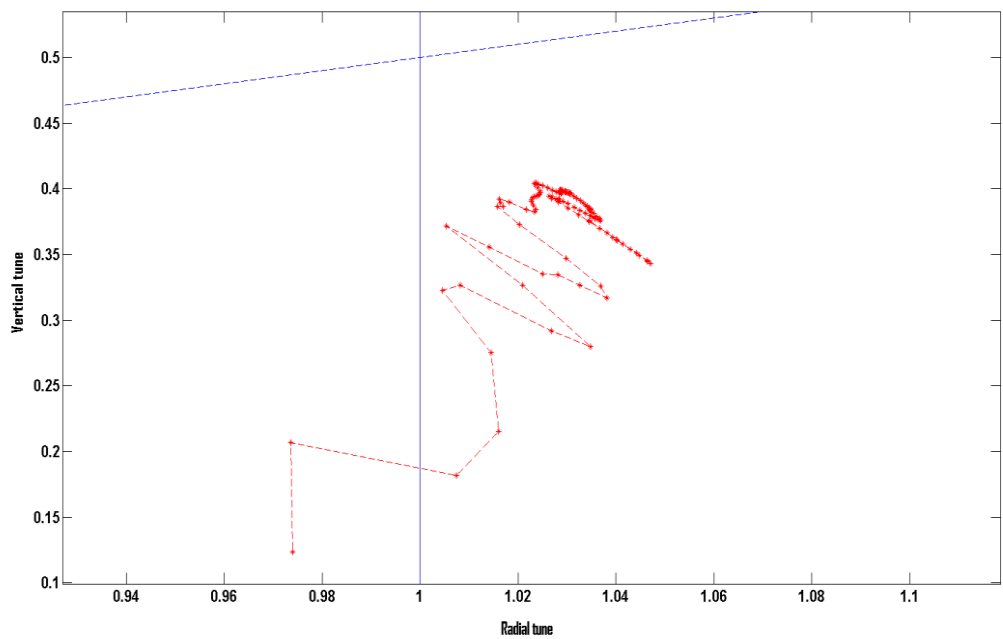

**Figure 2.** Structure-resonance diagram and excursions of the focusing frequencies from injection to extraction; (a) $2\nu_r + 2\nu_z = 4$, (b) $\nu_r - 2\nu_z = 0$, and (c) $4\nu_r = 4$.

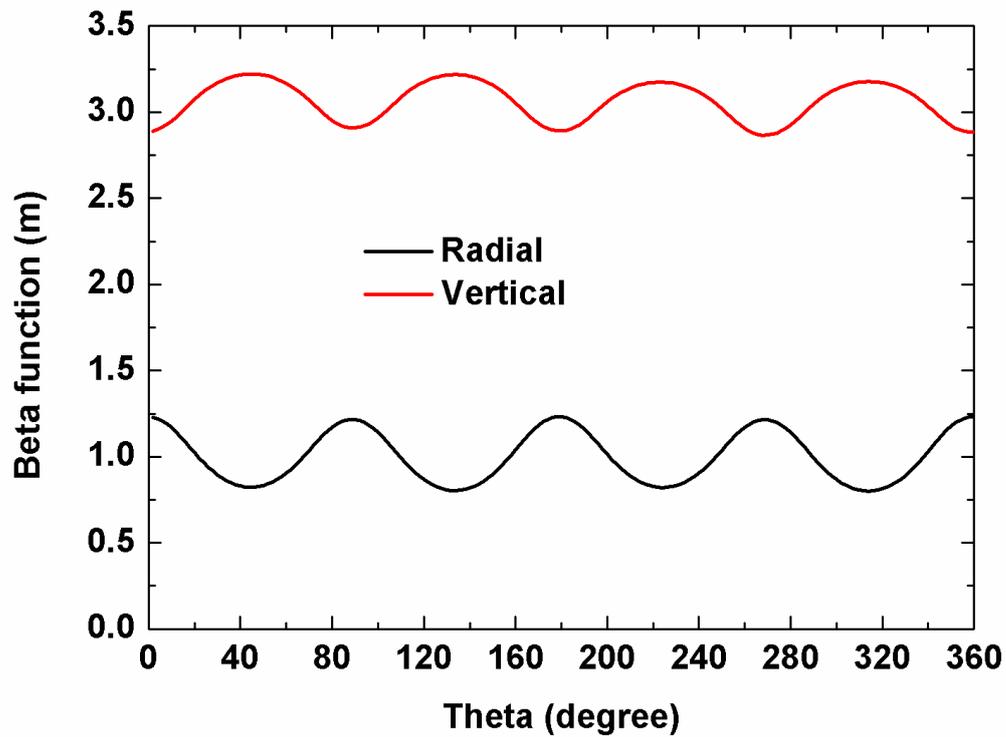

**Figure 3.** Determined beta-function along theta at 1 MeV. Beta-function shows 4 fold symmetry due to 4 sector cyclotron.

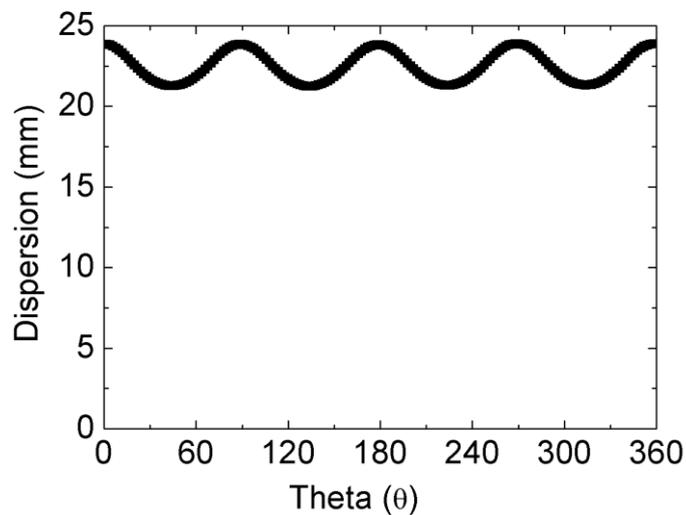

**Figure 4.** Determined dispersion-function along theta at 1 MeV. Dispersion-function also shows 4 fold symmetry due to 4 sector cyclotron.

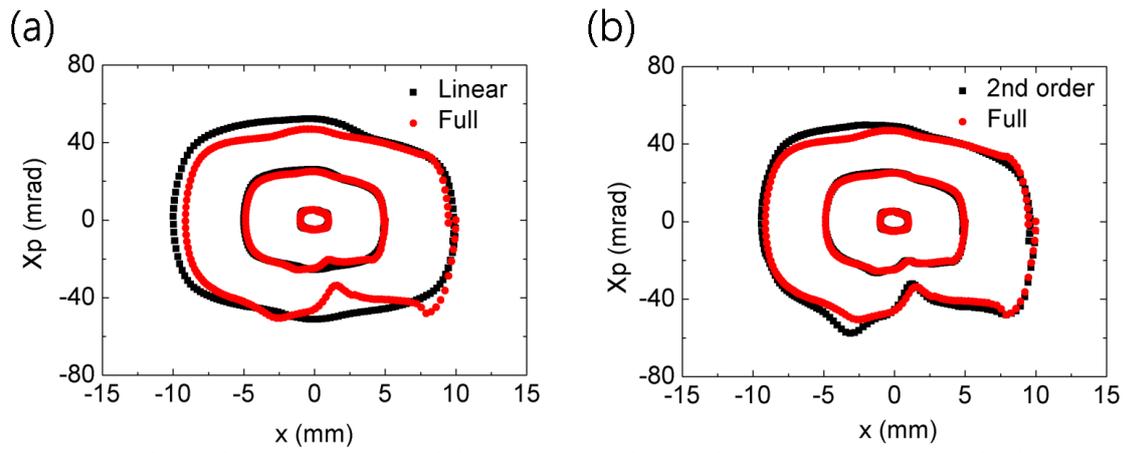

**Figure 5.** Horizontal phase-space. Red dots represent the result by Runge-Kutta method. Black dots represent phase space beam motion from (a) linear transfer matrix (b) the second order transfer matrix.

**Table 1.** Coefficient for the solution of inhomogeneous equation.

| Coefficient | Value |
|---|---|
| $f_{11}$ | $\frac{P^2}{p_\theta^3}X_{11}X_{21} + \frac{3}{2}\frac{rp_r P^2}{p_\theta^5}X_{21}^2$ |
| $f_{12}$ | $\frac{P^2}{p_\theta^3}(X_{11}X_{22} + X_{12}X_{21}) + 2 \times \frac{3}{2}\frac{rp_r P^2}{p_\theta^5}X_{21}X_{22}$ |
| $f_{13}$ | $\frac{P^2}{p_\theta^3}X_{12}X_{22} + \frac{3}{2}\frac{rp_r P^2}{p_\theta^5}$ |
| $f_{14}$ | $\frac{1}{2}\frac{rp_r}{p_\theta^3}Z_{21}^2$ |
| $f_{15}$ | $\frac{rp_r}{p_\theta^3}Z_{21}Z_{22}$ |
| $f_{16}$ | $\frac{1}{2}\frac{rp_r}{p_\theta^3}Z_{22}^2$ |
| $f_{21}$ | $\left(-\frac{\partial B}{\partial r} - \frac{r}{2}\frac{\partial^2 B}{\partial r^2}\right)X_{11}^2 - \frac{1}{2}\frac{P^2}{p_\theta^3}X_{21}^2$ |
| $f_{22}$ | $\left(-2\frac{\partial B}{\partial r} - r\frac{\partial^2 B}{\partial r^2}\right)X_{11}X_{12} - \frac{P^2}{p_\theta^3}X_{21}X_{22}$ |
| $f_{23}$ | $\left(-\frac{\partial B}{\partial r} - \frac{r}{2}\frac{\partial^2 B}{\partial r^2}\right)X_{12}^2 - \frac{1}{2}\frac{P^2}{p_\theta^3}X_{22}^2$ |
| $f_{24}$ | $\frac{1}{2}r\left(\frac{\partial^2 B}{\partial r^2} + \frac{1}{r}\frac{\partial B}{\partial r} + \frac{1}{r^2}\frac{\partial^2 B}{\partial \theta^2}\right)Z_{11}^2 + \frac{\partial B}{\partial \theta}\frac{1}{p_\theta}Z_{11}Z_{21} - \frac{1}{2}\frac{1}{p_\theta}Z_{21}^2$ |
| $f_{25}$ | $r\left(\frac{\partial^2 B}{\partial r^2} + \frac{1}{r}\frac{\partial B}{\partial r} + \frac{1}{r^2}\frac{\partial^2 B}{\partial \theta^2}\right)Z_{11}Z_{12} + \frac{\partial B}{\partial \theta}\frac{1}{p_\theta}(Z_{11}Z_{22} + Z_{12}Z_{21}) - \frac{1}{p_\theta}Z_{21}Z_{22}$ |
| $f_{26}$ | $\frac{1}{2}r\left(\frac{\partial^2 B}{\partial r^2} + \frac{1}{r}\frac{\partial B}{\partial r} + \frac{1}{r^2}\frac{\partial^2 B}{\partial \theta^2}\right)Z_{12}^2 + \frac{\partial B}{\partial \theta}\frac{1}{p_\theta}Z_{12}Z_{22} - \frac{1}{2}\frac{1}{p_\theta}Z_{22}^2$ |
| $g_{11}$ | $\frac{1}{p_\theta}X_{11}Z_{21} + \frac{rp_r}{p_\theta^3}X_{21}Z_{21}$ |
| $g_{12}$ | $\frac{1}{p_\theta}X_{11}Z_{22} + \frac{rp_r}{p_\theta^3}X_{21}Z_{22}$ |
| $g_{13}$ | $\frac{1}{p_\theta}X_{12}Z_{21} + \frac{rp_r}{p_\theta^3}X_{22}Z_{21}$ |
| $g_{14}$ | $\frac{1}{p_\theta}X_{12}Z_{22} + \frac{rp_r}{p_\theta^3}X_{22}Z_{22}$ |
| $g_{21}$ | $\left(\frac{\partial B}{\partial r} + r\frac{\partial^2 B}{\partial r^2} - \frac{p_r}{p_\theta}\frac{\partial^2 B}{\partial r\partial \theta}\right)X_{11}Z_{11} - \frac{p_r^2}{p_\theta^3}\frac{\partial B}{\partial \theta}X_{21}Z_{11}$ |
| $g_{12}$ | $\left(\frac{\partial B}{\partial r} + r\frac{\partial^2 B}{\partial r^2} - \frac{p_r}{p_\theta}\frac{\partial^2 B}{\partial r\partial \theta}\right)X_{11}Z_{12} - \frac{p_r^2}{p_\theta^3}\frac{\partial B}{\partial \theta}X_{21}Z_{11}$ |
| $g_{13}$ | $\left(\frac{\partial B}{\partial r} + r\frac{\partial^2 B}{\partial r^2} - \frac{p_r}{p_\theta}\frac{\partial^2 B}{\partial r\partial \theta}\right)X_{12}Z_{11} - \frac{p_r^2}{p_\theta^3}\frac{\partial B}{\partial \theta}X_{22}Z_{11}$ |
| $g_{14}$ | $\left(\frac{\partial B}{\partial r} + r\frac{\partial^2 B}{\partial r^2} - \frac{p_r}{p_\theta}\frac{\partial^2 B}{\partial r\partial \theta}\right)X_{12}Z_{22} - \frac{p_r^2}{p_\theta^3}\frac{\partial B}{\partial \theta}X_{22}Z_{12}$ |
| $h_1$ | $\frac{\gamma p_r}{p_\theta^3}X_{11}X_{21} + \frac{1}{2}\left(\frac{\gamma r}{p_\theta^3} + \frac{3\gamma r p_r^2}{p_\theta^5}\right)X_{21}^2$ |
| $h_2$ | $\frac{\gamma p_r}{p_\theta^3}(X_{11}X_{22} + X_{12}X_{21}) + 2 \times \frac{1}{2}\left(\frac{\gamma r}{p_\theta^3} + \frac{3\gamma r p_r^2}{p_\theta^5}\right)X_{21}X_{22}$ |
| $h_3$ | $\frac{\gamma p_r}{p_\theta^3}X_{12}X_{22} + \frac{1}{2}\left(\frac{\gamma r}{p_\theta^3} + \frac{3\gamma r p_r^2}{p_\theta^5}\right)X_{22}^2$ |
| $h_4$ | $\frac{1}{2}\frac{\gamma r}{p_\theta^3}$ |
| $h_5$ | $\frac{\gamma r}{p_\theta^3}$ |
| $h_6$ | $\frac{\gamma r}{p_\theta^3}$ |